\documentclass{emulateapj}
\usepackage{epsfig}
\begin{document}

\title{The relationship between mono-abundance and mono-age stellar populations in the Milky Way disk}
\author{I.~Minchev\altaffilmark{1}, M.~Steinmetz\altaffilmark{1}, C.~Chiappini\altaffilmark{1}, M.~Martig\altaffilmark{2}, F.~Anders\altaffilmark{1}, G. Matijevic\altaffilmark{1} and R.~S.~de~Jong\altaffilmark{1}
}

\altaffiltext{1}{Leibniz-Institut f\"{u}r Astrophysik Potsdam (AIP), An der Sternwarte 16, D-14482, Potsdam, Germany}
\altaffiltext{2}{Max-Planck-Institut f\"{ur} Astronomie, K\"{o}nigstuhl 17, D-69117 Heidelberg, Germany}

\begin{abstract}
Studying the Milky Way disk structure using stars in narrow bins of [Fe/H] and [$\alpha$/Fe] has recently been proposed as a powerful method to understand the Galactic thick and thin disk formation. It has been assumed so far that these mono-abundance populations (MAPs) are also coeval, or mono-age, populations. Here we study this relationship for a Milky Way chemo-dynamical model and show that equivalence between MAPs and mono-age populations exists only for the high-[$\alpha$/Fe] tail, where the chemical evolution curves of different Galactic radii are far apart. At lower [$\alpha$/Fe]-values a MAP is composed of stars with a range in ages, even for small observational uncertainties and a small MAP bin size. Due to the disk inside-out formation, for these MAPs younger stars are typically located at larger radii, which results in negative radial age gradients that can be as large as 2 Gyr/kpc. Positive radial age gradients can result for MAPs at the lowest [$\alpha$/Fe] and highest [Fe/H] end. Such variations with age prevent the simple interpretation of observations for which accurate ages are not available. Studying the variation with radius of the stellar surface density and scale-height in our model, we find good agreement to recent analyses of the APOGEE red-clump (RC) sample when 1-4 Gyr old stars dominate (as expected for the RC). Our results suggest that the APOGEE data are consistent with a Milky Way model for which mono-age populations flare for all ages. We propose observational tests for the validity of our predictions and argue that using accurate age measurements, such as from asteroseismology, is crucial for putting constraints on the Galactic formation and evolution.

\end{abstract}


\section{Introduction}

The formation of the Milky Way thin and thick disks is one of the most important topics in the area of Galactic Archaeology. The Galactic thick disk has been the subject of study since its discovery \citep{gilmore83, yoshii82}. Mechanisms of thick disk formation include vertical heating from infalling satellites (e.g., \citealt{villalobos08, quinn93}), turbulent gas-rich disk phase at high redshift (e.g., \citealt{bournaud09,forbes12}), massive gas-rich satellites \citep{brook04,brook05}, and accretion of satellite debris \citep{abadi03}.  

Formation of thick disks by radial migration was proposed as a mechanism by \cite{schonrich09b}, later on advocated by \cite{loebman11} and \cite{roskar13}. This idea was challenged by \cite{minchev11b,minchev12b}, demonstrating that migrators in N-body models do not have any significant effect on the disk thickening. Several independent groups have now supported these findings in more recent works \citep{martig14b, vera-ciro14, grand16, aumer16}, establishing this as a generic result of disk dynamics. The reason behind this is the conservation of vertical action of migrating populations (see \citealt{minchev11b, solway12, vera-ciro14}). Interestingly, when merger perturbations are accounted for, as in cosmological simulations, the effect of migration has a negative effect on the disk thickening (\citealt{mcm13}, hereafter MCM13; \citealt{mcm14}, hereafter MCM14; \citealt{grand16}).  

We recently proposed a new model for the formation of thick disks \citep{minchev15}. We showed that in galactic disks formed inside-out, mono-age populations (groups of coeval stars) are well fitted by single exponentials and always flare (the disk thickness increases with radius). In contrast, when the total stellar density is considered, a sum of two exponentials is required for a good fit, resulting in thin and thick disks which do not flare. We related this to the scale-length increase of younger populations, which flare at progressively larger radii. Such a scenario explains why chemically- or age-defined thick disks are centrally concentrated \citep{bensby11, bovy12a}, but geometrically thick populations in both observations of external edge-on galaxies \citep{yoachim06, pohlen07, comeron12} and in the Milky Way \citep{robin96, ojha01, juric08} extend beyond the thinner component. 

Flaring of mono-age disks also explains the inversion of metallicity gradients with increasing distance from the disk midplane, as younger metal-rich stars in the outer disk can reach high vertical distances. This inversion has been found in a number of spectroscopic Galactic surveys (e.g., SEGUE -- \citealt{cheng12a}, RAVE -- \citealt{boeche13b}, APOGEE -- \citealt{anders14}) and in simulations (MCM14, \citealt{minchev15, kawata16, miranda16}). A similar argument goes for the inversion of [$\alpha$/Fe] seen in observations \citep{boeche13b, anders14}. Both of the above phenomena result in models from a negative age gradient at high distance above the disk midplane. Such an age drop with radius was recently observed in the Milky Way \citep{martig16b} using APOGEE ages \citep{martig16a} estimated from the abundance ratio C/N and calibrated with asteroseismic Kepler data. 

The recent analyses of the APOGEE red-clump (RC) sample by \citealt{bovy16} (hereafter B16) suggest that the structure of groups of stars in narrow bins of [Fe/H] and [$\alpha$/Fe], known as mono-abundance populations (MAPs), is different for stars with high- and low-[$\alpha$/Fe] values. While the high-[$\alpha$/Fe] MAPs were found to have surface density profiles, $\Sigma(r)$, consistent with single exponentials in the range $4\leqslant r \leqslant14$~kpc and showed no flaring, the low-[$\alpha$/Fe] MAPs were exhibited peaks in $\Sigma(r)$ and did not flare. The lack of flaring in the high-[$\alpha$/Fe] MAPs was interpreted as evidence that the Milky Way thick disk was not created by the perturbative effect of mergers, which is expected to cause disk flaring in mono-age populations according to the analysis of \cite{minchev15}. The B16 interpretation assumes that a MAP is also a mono-age population.

In this paper we use the Milky Way chemo-dynamical model described in MCM13 and MCM14 (hereafter, the MCM model) to study the relationship between MAPs and mono-age populations and make a comparison to the results of B16. The disk structure of MAPs and mono-age populations in a disk formation simulation has been previously investigated by \cite{stinson13}, where it was concluded that MAPs are mostly comprised of co-eval stars. We will show that, while the latter is true, more significant differences between MAPs and mono-age populations become apparent when the variation of disk structure with galactic radius is considered. These can have important implications for the interpretation of observational data for which accurate stellar ages are not available.

\begin{figure}
\epsscale{1.2}
\plotone{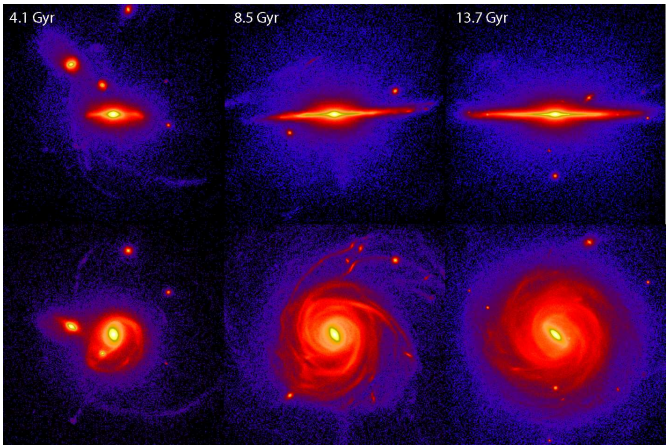}
\caption{
{\bf Top:} Edge-on view of our model galaxy at different time of the disk evolution. 
{\bf Bottom:} Face-on view of the disk at the same time outputs as above. An inside-out disk formation and strong perturbations by infalling satellites at high redshift is apparent from these snapshots.
}
\label{fig:xyz}
\end{figure}

\begin{figure}
\epsscale{1.2}
\plotone{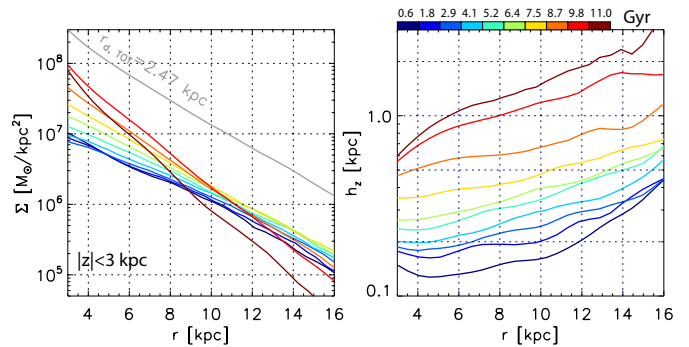}
\caption{
Disk stricture of model mono-age populations.
{\bf Left:} Stellar surface density, $\Sigma$, as a function of Galactocentric radius, $r$, for different age groups (color curves) using an age bin of one Gyr. More centrally concentrated older stellar populations are seen, consistent with an inside-out disk growth. The total disk scale-length in the shown radial range is indicated in gray.
{\bf Right:} Variation of stellar density scale-height, $h_z$, with $r$. Flaring is present for all age bins. 
}
\label{fig:hd}
\end{figure}

\begin{figure*}
\epsscale{1.15}
\plotone{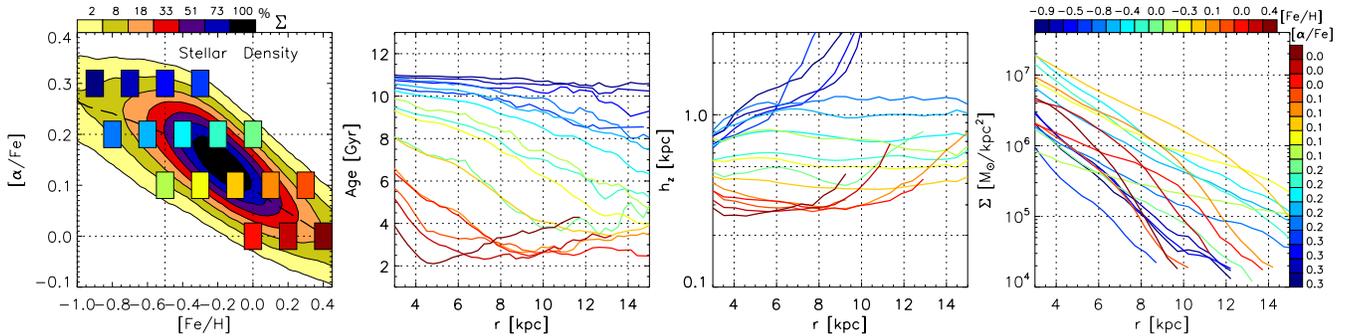}
\caption{
Disk stricture of model mono-abundance populations (MAPs).
{\bf Leftmost panel:} Stellar density contours for the model [$\alpha$/Fe]-[Fe/H] plane (contour levels shown on top). Overlaid are 17 MAPs of bin-sizes $\Delta[\alpha/\rm Fe]=0.05$ dex and $\Delta\rm[Fe/H]=0.1$ dex, centered on the values indicated by the two color bars in the rightmost panel.
{\bf Second panel:} Mean age variation with Galactocentric radius, $r$, for all MAPs shown in the leftmost panel using the same color coding. Only the most metal-poor and [$\alpha$/Fe]-rich MAPs are mono-age populations.
{\bf Third panel:} As second panel but vertical axis shows the stellar density scale-height, $h_z$.  Strong flaring is seen only when mean age is constant with radius.
{\bf Rightmost panel:} As second and third panels but showing the stellar surface density, $\Sigma$, as a function of $r$. $\Sigma$ falls off as a single exponential for the high-[$\alpha$/Fe] MAPs (blueish colors) but type-II breaks are seen for MAPs at lower [$\alpha$/Fe] values.
}
\label{fig:all}
\end{figure*}

\section{The model data}
\label{sec:chem}

To properly model the Milky Way it is crucial to be consistent with observational constraints at redshift $z=0$, for example, a flat rotation curve, a small bulge, a central bar of an intermediate size, gas to total disc mass ratio of $\sim0.14$ in the solar vicinity, and local disc velocity dispersions close to the observed ones.

While cosmological simulations would be the natural framework for a state-of-the-art chemo-dynamical study of the Milky Way, a number of star formation and chemical enrichment problems still exist in fully self-consistent simulations (see discussion in MCM13). To avoid these problems, MCM13 created a hybrid chemo-dynamical model using a high-resolution simulation in the cosmological context coupled with a pure chemical evolution model.

The simulation used for the MCM model is part of a suite of numerical experiments presented by \cite{martig12}, where the authors studied the evolution of 33 simulated galaxies from $z=5$ to $z=0$ using the zoom-in technique described by \cite{martig09}. This technique consists of extracting merger and accretion histories for a given halo in a $\Lambda$CDM cosmological simulation and then re-simulating at much higher resolution (150~pc spatial, and 10$^{4-5}$~M$_{\odot}$ mass resolution). 

The general formation and evolutionary behavior of the disk formation is similar to many recent simulations in the cosmological context (e.g., \citealt{brook12, stinson13, bird13, marinacci14}). An initial central component is formed during an early epoch of violent merger activity, where gas-rich mergers supply the initial reservoir of gas at high redshift and merger activity decreases with redshift, similarly to what is expected for the Milky Way. This inside-out disk formation results in older stellar populations being centrally concentrated (see Fig.~\ref{fig:xyz} and left panel of Fig.~\ref{fig:hd}). The right panel of Fig.~\ref{fig:hd} shows disk flaring for all mono-age groups considered.

The chemical evolution model (CEM) used for the MCM chemo-dynamical model was similar to that of \cite{chiappini09a} and is described in MCM13. The disks grow inside-out in both the CEM and the simulation, although not exactly at the same rate due to the unconstrained nature of the simulation. A comparison between the two star formation histories (SFHs) as a function of cosmic time is shown in Fig.~A.1 by MCM14. To be consistent with the CEM chemical enrichment, which is the direct result of the SFH, the simulation SFH was weighted to match that of the CEM. Further details about the MCM model, comparison to observations and prediction for future surveys can be found in MCM13, MCM14, \cite{minchev14, minchev16a, anders16a, anders16b}.

\begin{figure}
\epsscale{1.15}
\plotone{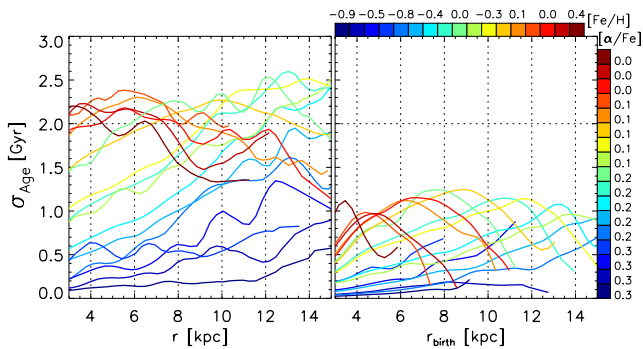}
\caption{
Age dispersion, $\sigma_{Age}$, as a function of final (left) and initial (right) radius for the model MAPs. Comparing to the second panel of Fig.~\ref{fig:all}, we find inverse correlations with the mean age radial trend for all MAPs. Very low age dispersion is found for the highest-[$\alpha$/Fe] MAPs, while the largest spread is in the most metal-rich and [$\alpha$/Fe]-poor stars. The contrast between $\sigma_{Age}(r)$ and $\sigma_{Age}(r_{birth})$ indicates the effect of migration -- stars with a large spread in birth radii and with a range in ages can have similar chemical compositions (i.e., fall in the same MAP), thus increasing the age scatter at a given final radius.
}
\label{fig:sig_age}
\end{figure}

For comparison with the APOGEE data we constrain spatially our model in the range $3\leqslant r\leqslant15$ kpc and $|z|\leqslant3$ kpc. As in the data, we define the ratio $[\alpha/\rm Fe]=[(\rm [O+Mg+Si+S+Ca]/5)/Fe]$ using the model's chemical abundances.

B16 estimated random uncertainties of  $\delta[\alpha/\rm Fe]=0.02$ dex and $\delta\rm[Fe/H]=0.05$ dex, empirically determined using scatter in open clusters. To make our results comparable to the data, we implement uncertainties in our model [$\alpha$/Fe] and [Fe/H] drawing from Gaussian distributions with standard deviations 0.02 dex and 0.05 dex, respectively. Also as in B16, to define a MAP we use bins of $\Delta[\alpha/\rm Fe]=0.05$ dex and $\Delta\rm[Fe/H]=0.1$ dex. B16 calibrated the APOGEE chemical abundances to the solar abundance scale using the open cluster M67 by applying constant offsets of $-0.1$ dex and $-0.05$ dex to the APOGEE abundance scales of [Fe/H] and [$\alpha$/Fe], respectively. To normalize our model to the B16 APOGEE RC-sample, in Fig.\ref{fig:match} we apply an offset of $-0.1$ dex in [Fe/H].

\section{Results}

\subsection{When is a MAP a mono-age population?}
\label{sec:def}

One way to answer the above question is by considering the age-[Fe/H] and age-[$\alpha$/Fe] relations in a Milky Way chemical evolution model.  A MAP would be a mono-age population only if its constituent stars have a spread in age no larger than that expected from the MAP bin size used. 

In our chemical evolution model, for $r\geqslant4$ kpc and look-back-time $\leqslant8-9$ Gyr the interstellar medium temporal increase in [Fe/H] and decrease in [$\alpha$/Fe] is $\sim0.05$ dex/Gyr and $\sim0.03$ dex/Gyr, respectively (see Figures~4 and 7 in MCM13). Considering the MAP bin size of $\Delta\rm[Fe/H]=0.1$ dex and $\Delta[\alpha/\rm Fe]=0.05$ dex used by B16 and this work, in order for a MAP to be a mono-age population it should consist of stars that span an age range of no more than two Gyr. 

For the old, high-[$\alpha$/Fe] stellar populations the chemical evolution is much faster: in the period 9-11 Gyr at $r=8$ kpc the rate of change in [Fe/H] and  [$\alpha$/Fe]  is $\sim0.3$ dex/Gyr and $\sim0.1$ dex/Gyr, respectively. In this case, for the MAP bin size used in this work, a mono-age population needs to have an age range of no more than $\sim0.3-0.5$ Gyr. For the discussion in this paper, keeping in mind the difficulty of measuring ages with a Gyr precision for 10-Gyr-old stars, we will consider two Gyr to be sufficiently small to define a mono-age population.

\subsection{Mean age variation with radius of model MAPs}
\label{sec:mean}

The leftmost panel of Fig.~\ref{fig:all} shows stellar density contours for the model [$\alpha$/Fe]-[Fe/H] plane. Overlaid are 17 MAPs of bin-sizes $\Delta[\alpha/\rm Fe]=0.05$ dex and $\Delta\rm[Fe/H]=0.1$ dex, centered on the [$\alpha$/Fe] and [Fe/H] values indicated by the two color bars in the rightmost panel.

The second panel of Fig.\ref{fig:all} shows the mean age variation with Galactocentric radius, $r$, for each MAP shown in the leftmost panel. The same color-coding is used. An interesting observation is that only the most metal-poor and [$\alpha$/Fe]-rich MAPs show no age variation over the radial range shown, i.e., can be considered mono-age populations. Even for the highest [$\alpha$/Fe] MAPs, as [Fe/H] increases to -0.5 and then -0.3 dex (top row of bins in leftmost panel) a drop in mean age is seen at $r>12$ and $r>8$ kpc, respectively. The clustering of the highest [$\alpha$/Fe] MAPs at $\sim10-11$~Gyr is related to the fact that, for our MAP bin size, a MAP at these regions of the [$\alpha$/Fe]-[Fe/H] plane corresponds to a narrow range in age (see \S\ref{sec:def}). Larger separation among the MAPs mean age is seen for lower-[$\alpha$/Fe] and higher-[Fe/H] MAPs because the star formation rate decreases with time. This allows the same MAP bin size to host stars with increasingly larger age range. As [$\alpha$/Fe] drops to 0.2 dex (second row of bins in leftmost panel of Fig.\ref{fig:all}) the negative radial age gradient grows to $\sim2$ Gyr/kpc. Finally, at [$\alpha$/Fe]$<0.05$ and [Fe/H]$>0.0$ dex a flattening and inversion in mean age at larger radii results.The radius at which this takes place increases from $\sim5$ kpc to $\sim12$ kpc as MAP position changes from for ([$\alpha$/Fe], [Fe/H])$=(0.0, 0.4)$ to $(0.1, -0.3)$. 

The third panel of Fig.~\ref{fig:all} shows the radial variation of the stellar density scale-height, $h_z$.  Similarly to the mono-age populations studied in \cite{minchev15}, single exponential fits are found to be sufficient for all MAPs.  Strong flaring is seen only when the mean age is relatively constant with radius, as is the case for the MAPs at [$\alpha$/Fe]$=0.3$ dex. For the MAPs at [$\alpha$/Fe]$=0.2$ dex $h_z(r)$ is mostly flat. This is related to the negative age gradients seen in the second panel of the figure. Flaring reappears for MAPs at [$\alpha$/Fe]$<0.05$ and [Fe/H]$>0.0$, which is where the age variation with radius flattens and inverts.

The rightmost panel of Fig.~\ref{fig:all} shows the stellar surface density, $\Sigma$, as a function of $r$ for the same MAPs as in the previous panels. $\Sigma(r)$ falls off as a single exponential for the high-[$\alpha$/Fe] MAPs (blueish colors) but type-II breaks appear for MAPs at lower [$\alpha$/Fe] values. This plot is very similar to the second and third top panels of Fig.~11 by MCM14, where it was already noted that the lack of good fit by single exponential profiles for low-[$\alpha$/Fe] high-[Fe/H] stars was in contrast to the results of \cite{bovy12a} using the SEGUE G-dwarf sample. 

\begin{figure*}
\epsscale{1.15}
\plotone{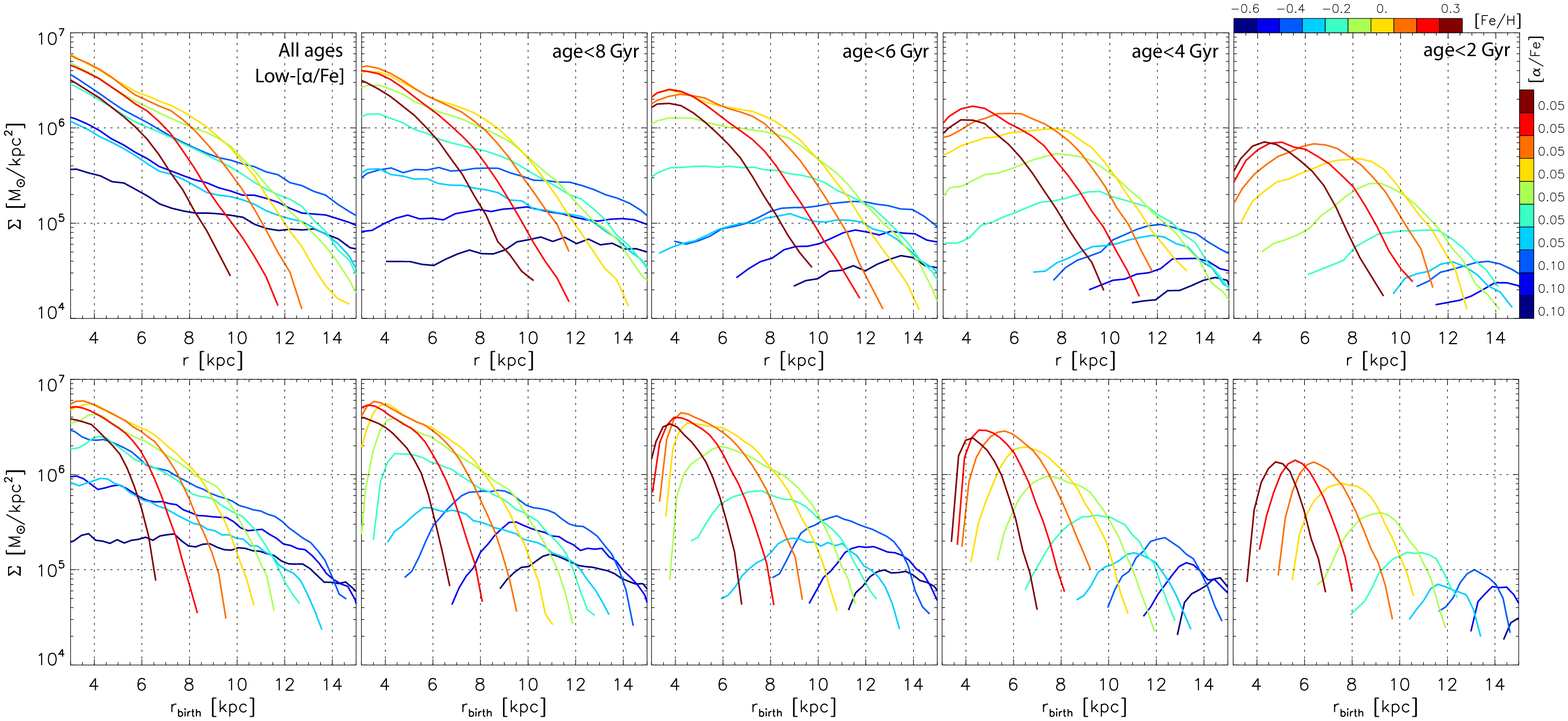}
\caption{
{\bf Top:} Surface density profiles of low-[$\alpha$/Fe] MAPs for the MCM model. The [$\alpha$/Fe] and [Fe/H] bins are the same as the low-[$\alpha$/Fe] MAPs in B16. In the left-hand panel we show stars of all ages, while in the other panels stars become progressively younger, as indicated. As older stars are discarded, all MAPs start to exhibit peaks, which shift to larger radii the more metal-poor the MAP, similarly to the APOGEE RC-sample. 
{\bf Bottom:} Same as top but plotted versus birth-radius.
}
\label{fig:age}
\end{figure*}

\begin{figure*}
\epsscale{1.15}
\plotone{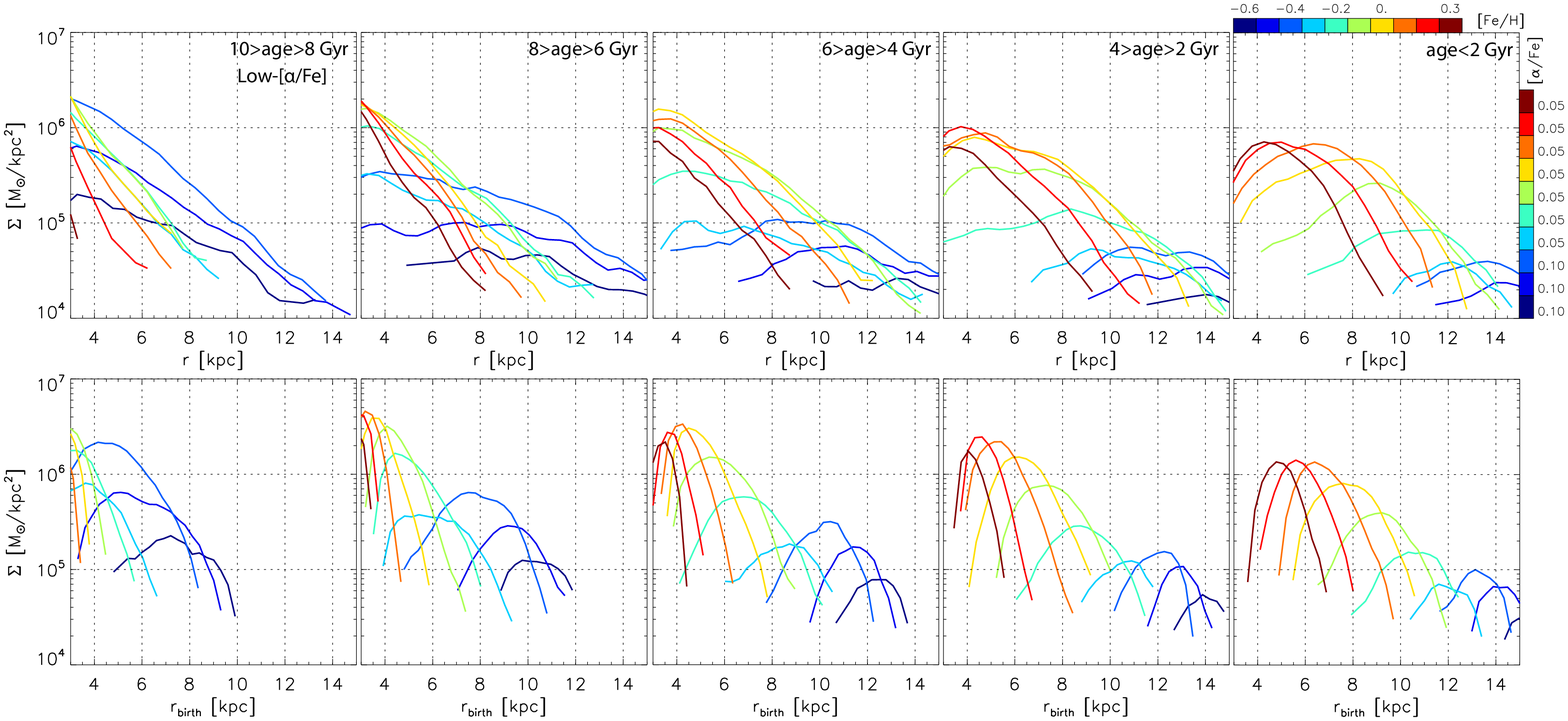}
\caption{
Same as Fig.~\ref{fig:age} but for groups of stars of common age, i.e., mono-age populations, as indicated. As the mean age decreases the density peaks shift to larger birth radii for all MAPs. This explains the wide birth radius distributions at the cumulative age cuts in Fig.~\ref{fig:age}. Because MAPs can be composed of a range of ages, to put constraints on the Galactic chemo-dynamical evolution it is needed to decompose MAPs in mono-age populations.  
}
\label{fig:age1}
\end{figure*}

\subsection{Are MAPs mono-age populations at a fixed radius?}

We have just demonstrated that, for our model, MAPs are not mono-age population, except at the highest [$\alpha$/Fe] values. We now study the variation with radius of the age dispersion, which will give us an indication of whether MAPs are mono-age populations for stars at a fixed galactic radius.

Fig.~\ref{fig:sig_age} shows the age dispersion, $\sigma_{Age}$, as a function of final (left panel) and birth (right panel) radius for the model MAPs. At each radius $\sigma_{Age}$ is estimated as the standard deviation of stars in a radial bin of 0.5 kpc. A first impression from looking at this figure is the overall increase of $\sigma_{Age}$ with decreasing [$\alpha$/Fe], which is an indication of the degeneracy introduced by the closely spaced chemical evolution curves of different Galactic radii at low [$\alpha$/Fe] (see bottom left panel in Fig.4 by MCM13).

Contrasting $\sigma_{Age}(r)$ with $Age(r)$ (second panel of Fig.~\ref{fig:all}) we find an inverse correlation in the radial trends of most MAPs. Very low age dispersion exists for the highest-[$\alpha$/Fe] MAPs, while the largest age spread is in the most metal-rich and [$\alpha$/Fe]-poor stars.

A significant decrease in $\sigma_{Age}$ is seen for all MAPs and radii when plotted against the mean birth radius (right panel of Fig.~\ref{fig:sig_age}). This contrast between $\sigma_{Age}(r)$ and $\sigma_{Age}(r_{birth})$ is indicative of the effect of radial migration -- stars with a large spread in birth radii and with a range in ages can have similar chemical compositions (i.e., fall in the same MAP), thus increasing the age scatter at a given final radius. 

According to our definition in \S\ref{sec:def}, all MAPs of non-migrators are mono-age populations. Assuming a Gaussian age distribution, the maximum age dispersion of one Gyr seen in the right panel of Fig.~\ref{fig:sig_age} then corresponds to an age range of about two Gyr. As migration takes place $\sigma_{Age}$ more than doubles (left panel of Fig.~\ref{fig:sig_age}). This, however, has no large effect on the highest-[$\alpha$/Fe] MAPs, which is consistent with the conclusion from \S\ref{sec:mean} that these MAPs are the only mono-age populations.
 
For MAPs centered on intermediate metallicity and [$\alpha$/Fe] values (the majority of stars) there is a prominent positive gradient of $\sim1.5$ Gyr/kpc in age dispersion. This trend in $\sigma_{Age}(r)$ is mostly due to the increase with radius in the fraction of migrators to non-migrators described and discussed by MCM14 using the same model we use here, although some of these variations are already seen in  $\sigma_{Age}(r_{birth})$.

Oscillations in $\sigma_{Age}(r)$ with a wavelength of about two kpc are seen in many of the MAPs in Fig.~\ref{fig:sig_age}. This is consistent with the expected typical jumps stars execute during migration (e.g., \citealt{roskar12}). The structure in $\sigma_{Age}(r)$, therefore, may be possible to relate to the radial migration efficiency as a function of time and galactic radius.

\subsection{Variation with age of model MAPs surface density profiles}

We showed in the previous sections that our model MAPs can be composed of stars with a range of ages. It is therefore interesting to explore how a MAP scale-length changes when the age range changes. For the sake of comparison to the APOGEE data we pick MAPs with the same central [$\alpha$/Fe] and [Fe/H] values as the low-[$\alpha$/Fe] MAPs defined by B16. We remind the reader that for the comparison with the APOGEE RC-sample we apply an offset of $-0.1$ dex in [Fe/H].

In the top row of Fig.~\ref{fig:age} we plot the surface density as a function of Galactic radius, $\Sigma(r)$, for our model low-[$\alpha$/Fe] MAPs. The leftmost panel shows stars of all ages; moving rightward, stars become progressively younger, as indicated. When stars of all ages are considered, breaks in $\Sigma(r)$ exist for MAPs with [Fe/H]$\leqslant-0.1$ dex and the decrease in density is always monotonic. This is in contrast to the APOGEE RC-sample, where all low-[$\alpha$/Fe] MAPs were found to exhibit peaks. 
As older stars are discarded, however, $\Sigma(r)$ starts to decrease at smaller radii due to the more centrally concentrated older populations. For the subsample with age $<4$ Gyr we find peaks in $\Sigma(r)$ for all low-[$\alpha$/Fe] MAPs, which is very similar to the results of B16. This result is consistent with the expectation that the APOGEE RC-sample peaks in age at 1-4 Gyr (B16, \citealt{girardi16}).

The bottom row of Fig.~\ref{fig:age} is similar to the top one, but $\Sigma$ is plotted versus the birth-radius, $r_{birth}$. The differences between the shape of $\Sigma(r)$ and $\Sigma(r_{birth})$ gives an indication of how much radial migration has taken place in the model. For all age subsamples MAPs are much more concentrated at birth (bottom) compared to the final time (top). More metal-poor populations always peak further out in the disk as expected for an inside-out formation and as seen in APOGEE.

To explore further, in Fig.~\ref{fig:age1} we show again $\Sigma(r)$ and $\Sigma(r_{birth})$ for all low-[$\alpha$/Fe] MAPs, but this time mono-age populations of width 2 Gyr are considered. 
It is clear from this figure that the $\Sigma(r)$ peaks shift to lower radii as the sample age increases. This is especially obvious for $\Sigma(r_{birth})$. For example, for the MAP centered on ([$\alpha$/Fe],Fe/H)=(0.1,-0.5) the peak shifts from $\sim14$ to $\sim7$ kpc as age increases from the youngest to the oldest age bin shown in the figure. This overall peak shift to lower radii as age increases explains the wider $\Sigma(r_{birth})$ profiles in the bottom row of Fig.~\ref{fig:age}.

\subsection{Matching the APOGEE RC-sample}

\begin{figure}
\epsscale{1.2}
\plotone{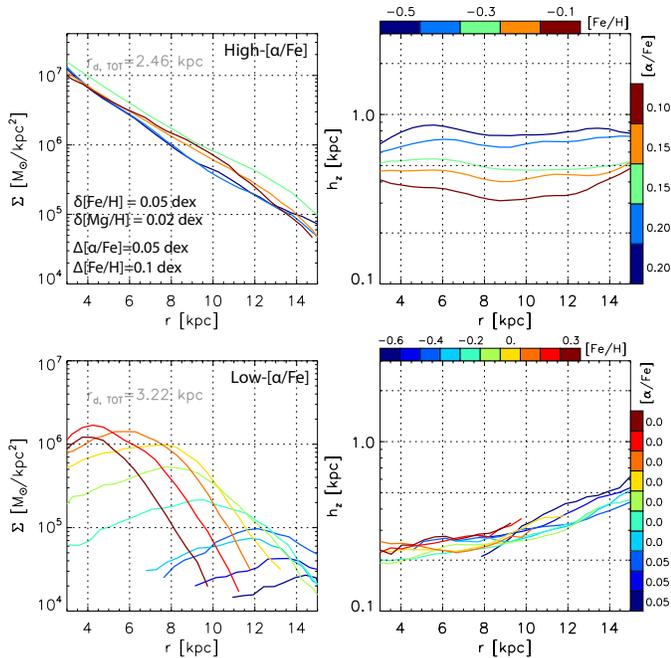}
\caption{
{\bf Left column}: A match to Fig.~11 by B16. As in the APOGEE-RC sample, the high-[$\alpha$/Fe] MAPs (in this case [$\alpha$/Fe]) do not display breaks or peaks in their surface-density profiles but are consistent with single exponentials. Also in agreement with the observations, the low-[$\alpha$/Fe] MAPs show peaks at increasingly larger radii the lower the [Fe/H] bin. 
{\bf Right column}: A match to Fig.~13 by B16. As in the APOGEE-RC sample, the high-[$\alpha$/Fe] MAPs do not show any flaring, i.e., their scale-height, $h_z$, does not increase with radius. In contrast, the low-[$\alpha$/Fe] MAPs flare. 
}
\label{fig:match}
\end{figure}

As discussed above, to match the surface density profiles of the APOGEE low-[$\alpha$/Fe] MAPs we need a sample peaking at low ages (Fig.~\ref{fig:age}). For the high-[$\alpha$/Fe] MAPs, however, single exponential $\Sigma(r)$ are well reproduced by considering the model unbiased sample. This is because the RC-sample bias toward younger stars is not important at high [$\alpha$/Fe]. The model density profiles for the exact APOGEE MAPs considered by B16 are presented in the left column of Fig.~\ref{fig:match}. 

It was already shown in Fig.~\ref{fig:all} that the disk scale-heights for MAPs at [$\alpha$/Fe]$\leqslant0.2$ dex are flat with radius. This lack of flaring in a model for which all mono-age populations flare is due to the negative age gradient present for MAPs in this region of the [$\alpha$/Fe]-[Fe/H] plane. The highest [$\alpha$/Fe] MAPs considered by B16 are centered on [$\alpha$/Fe]$=0.2$ dex, which explains the flat scale-height profiles they found. In Fig.~\ref{fig:all} it was also seen that for MAPs centered on [$\alpha$/Fe]$=0.0$ dex flaring is reappearing due to the inversion in the negative age gradient at this region of the [$\alpha$/Fe]-[Fe/H] plane. This is also true for MAPs at the slightly higher values of [$\alpha$/Fe]$=0.05$ dex. The model scale-heights for the exact APOGEE MAPs considered by B16 are presented in the right column of Fig.~\ref{fig:match}. 

We note that, in view of our findings, neither the low- nor the high-[$\alpha$/Fe] MAPs, shown in Fig.~\ref{fig:match}, are directly indicative of the Milky Way disk evolution because of the range in ages found in each MAP. Age information is necessary to assess the disk chemo-dynamical evolution.

\subsection{The effect of MAP bin size}

\begin{figure}
\epsscale{1.2}
\plotone{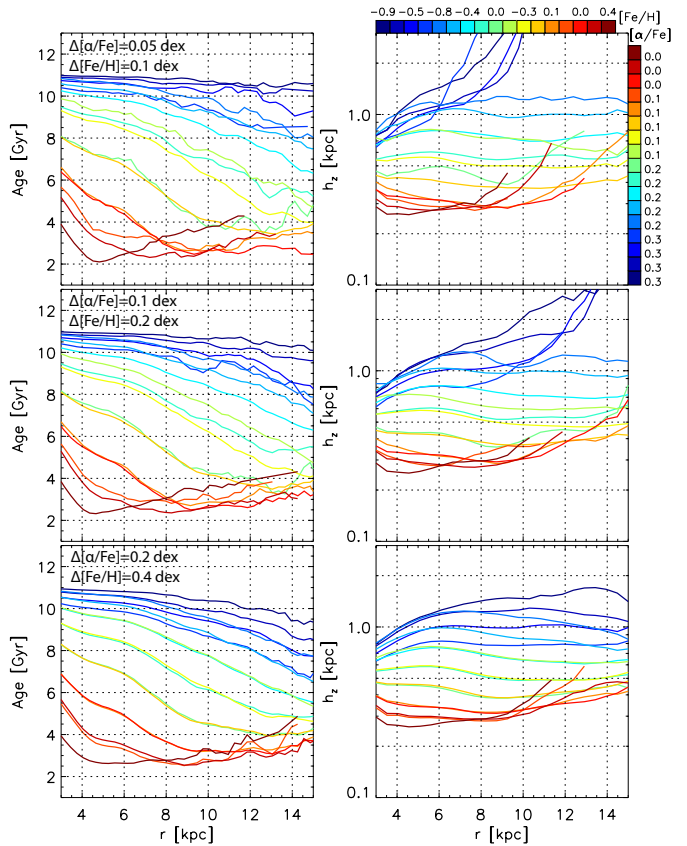}
\caption{
Assessing the effect of a MAP bins size. {\bf Top row}: Identical to the second and third top panels of Fig.~\ref{fig:all}. The [$\alpha$/Fe] and [Fe/H] bins sizes are as in B16. 
{\bf Middle row}: An increase in the bin sizes by a factor of two, as indicated. The strong flaring in for the highest-[$\alpha$/Fe] MAPs is decreased significantly. 
{\bf Bottom row}: As the bins sizes are increased by a factor of four the flaring almost completely disappears
}
\label{fig:bins}
\end{figure}

We assess in Fig.~\ref{fig:bins}  how our results are affected by increasing the MAP bin size. The top row of the figure shows age$(r)$ and $h_z(r)$ for 17 model MAPs and is identical to the corresponding panels of Fig.~\ref{fig:all}. The second row shows the same plots but the [$\alpha$/Fe] and [Fe/H] bin sizes are increased by a factor of two. The most obvious effect of this is the decrease in disk flaring for the highest [$\alpha$/Fe] MAPs. As the MAP bin size increases by a factor of four (bottom row of Fig.~\ref{fig:bins}), the flaring almost completely disappears, except for the most metal-poor highest-[$\alpha$/Fe] bin. We can see from the left row that the disk flaring is removed because of the negative radial age gradients resulting from the contamination by younger, outer disk stars, as the bin size increases. Some decrease in flaring is also seen for the most metal-rich and [$\alpha$/Fe]-poor MAPs.

Fig.~\ref{fig:bins} showed that increasing the MAP bin size turns the highest-[$\alpha$/Fe] flat radial age gradients negative and, thus, removes their flaring. The contamination of MAPs by neighboring bins is a function of position in the [$\alpha$/Fe]-[Fe/H] plane, the MAP bin size, and the abundance uncertainties. The uncertainties of 0.02 and 0.05 dex in [$\alpha$/Fe]  and [Fe/H], respectively, quoted by B16 and convolved into our model are at the highest level of precision we can hope Galactic spectroscopic surveys can achieve. Therefore, according to our model, except at the highest [$\alpha$/Fe] of 0.3 dex, we cannot hope to find MAPs which are also mono-age population.  

While at high-[$\alpha$/Fe] a small bin size and small uncertainties can result in MAPs being mono-age populations (i.e., no age variation with radius), this is not possible at low [$\alpha$/Fe], due to the age-abundance degeneracy expected in this part of the [$\alpha$/Fe]-[Fe/H] plane. 

\section{Discussion and conclusions}

We used a chemo-dynamical model tailored to the Milky Way and consistent with a number of observational constraints, to study the relation between MAPs and mono-age populations in the radial range $3\leqslant r\leqslant15$ kpc. We found that only MAPs at the highest-[$\alpha$/Fe] values are also mono-age populations (Fig.~\ref{fig:all}). At [$\alpha$/Fe]$<0.25$ dex a MAP is composed of stars with a range of ages, despite the small errors implemented and the small bin size. Due to the inside-out disk formation, younger stars are typically deposited at larger radii, which gives rise to a negative age gradient for a given MAP. 

Therefore, although the model disk flares for all mono-age populations (Fig.~\ref{fig:hd}, right), flaring representative of a mono-age population is found only for MAPs at [$\alpha$/Fe]$\geqslant0.25$ (Fig.~\ref{fig:all}). For MAPs at $0.05\leqslant$[$\alpha$/Fe]$\leqslant0.25$ dex the resulting negative age gradients cause a relatively flat scale-height variation with radius. 
 
At [$\alpha$/Fe]$<0.05$ dex an inversion from a negative to a positive slope in age is found for a given MAP, where the turning point shifts to larger radii as metallicity decreases. Here flaring is also found, but it is not characteristic of that in a mono-age population, i.e., we cannot use this to directly draw conclusions about the disk evolution as a MAP here represents a mix of populations.

The age dependence of MAPs is also reflected in the shape of the surface density profiles, $\Sigma(r)$, where, as age decreases, peaks in $\Sigma(r)$ form for all MAPs at low-[$\alpha$/Fe] (Fig.~\ref{fig:age})

We demonstrated that our model can match well the surface density and scale-height radial profiles of MAPs in the APOGEE RC-sample. For model MAPs consistent with the high-[$\alpha$/Fe] APOGEE MAPs studied by B16 we found no flaring, as in the data. The reason for this is that the highest [$\alpha$/Fe] MAPs B16 considered were centered on [$\alpha$/Fe]$=0.2$ dex, which is in the region where age mixing occurs (see Fig.~\ref{fig:all}). MAP structure could not be explored at higher [$\alpha$/Fe] and [Fe/H]$<-0.5$ dex due to the low statistics of the APOGEE RC-sample in that region (see Fig.~5 by B16). The upcoming APOGEE data release 14 in 2017 will increase the RC-sample by about 60\%. This will make it possible to test our prediction that MAPs should show disk flaring at [$\alpha$/Fe]$>0.25$ dex and even possibly at the currently explored [$\alpha$/Fe]=0.2 dex but at metallicities lower than -0.5 dex. 

We note that depending on the $\alpha$-element used, e.g., O, Mg, Si, S, Ca, or a combination thereof, as in B16 and this work, the range of [X/Fe] (where X is an $\alpha$-element) will be different. For example, in the case of O an equivalence between MAPs and mono-age populations would most likely happen at a higher value, i.e., at [O/Fe]$>0.4$. Future work should explore the relation between MAPs and mono-age populations for different individual elements and different Galactic models.

For the high-[$\alpha$/Fe] model MAPs we found surface density profiles consistent with single exponentials, as in B16. We could also explain the peaks of the low-[$\alpha$/Fe] MAPs density profiles as the result of a bias toward younger ages in the APOGEE RC-sample. 
The $\Sigma(r)$ peak location is a function of position in the [$\alpha$/Fe]-[Fe/H] plane and related to the radial and temporal variation of metallicity in the disk. The fact that the peak shifts outwards as metallicity decreases indicates an inside-out disk formation.  The peak amplitude is meaningful only in a bias-corrected observational sample. The $\Sigma(r)$ spread around the peak of a given MAP is an indication of the heating and angular momentum redistribution that have taken place in the disk. 

Another testable prediction using chemo-kinematical information only is that the peaked surface density profiles found for the low-[$\alpha$/Fe] MAPs of the APOGEE RC-sample should disappear as a sample representing better the disk older stellar population in the Galaxy is studied. This is because for each MAP older stars will be preferentially concentrated in the inner disk and will thus change the shape of $\Sigma(r)$ mostly inwards of the density peak. 

Alternatively, splitting the RC-sample into mono-age populations (even as wide as two Gyr) should reveal a shift in the MAP density peaks to lower radii, the older the age group, while eventually losing the peak for the oldest age groups (see Fig.~\ref{fig:age1}). 

Radial migration is a strong function of time and Galactic radius with especially active radii at the bar's CR and 2:1 OLR (e.g., \citealt{mf10, brunetti11}). It may be possible to constrain this by searching for variations in the surface density radial profiles of mono-age populations for a given MAP. For example, in Fig.~\ref{fig:age1} we find mostly smooth single peaks for $\Sigma(r_{birth})$ for MAPs of different ages (bottom row) but as migration takes place wiggles in the final $\Sigma(r)$ can be seen (top row). These density fluctuations are washed out for the oldest stars because of their larger velocity dispersions. It may be very challenging, but not impossible, to relate this structure in $\Sigma(r)$ to the migration efficiency as a function of time and Galactic radius, as age uncertainties improve in the near future.

Another way of constraining the migration temporal and spatial evolution may be by relating the fluctuations in the age dispersion radial profiles of MAPs (Fig.~\ref{fig:sig_age}) to resonant locations associated with dynamical instabilities.

It is clear from the above discussion that age information is very important for recovering the disk chemo-dynamical evolution. Asteroseismology will play a key role in this, as it gives the opportunity to obtain ages over large distances. It is seen from our results that age precision of around one Gyr (half the age bin size used in Fig.~\ref{fig:age1}) would be enough to test our predictions. This seems to be already feasible (e.g., \citealt{anders16a, noels15}), although larger samples with seismic information, covering several lines of sight are necessary, as in a combination of all K2 and CoRoT fields. 

\acknowledgments
We thank the anonymous referee for a constructive report that helped improve the manuscript. IM acknowledges support by the Deutsche Forschungsgemeinschaft under the grant MI 2009/1-1.


\begin{thebibliography}{}

\bibitem[\protect\astroncite{{Abadi} et~al.}{2003}]{abadi03}
{Abadi}, M.~G., {Navarro}, J.~F., {Steinmetz}, M., and {Eke}, V.~R.: 2003,
\newblock {\em \apj} {\bf 597}, 21

\bibitem[\protect\astroncite{{Anders} et~al.}{2016a}]{anders16a}
{Anders}, F., {Chiappini}, C., {Rodrigues}, et al.: 2016a,
\newblock {\em ArXiv e-prints}

\bibitem[\protect\astroncite{{Anders} et~al.}{2016b}]{anders16b}
{Anders}, F., {Chiappini}, C., {Rodrigues}, T.~S., {Piffl}, T., {Mosser}, B.,
  {Miglio}, A., {Montalb{\'a}n}, J., {Girardi}, L., {Minchev}, I., {Valentini},
  M., and {Steinmetz}, M.: 2016b,
\newblock {\em ArXiv e-prints}

\bibitem[\protect\astroncite{{Anders} et~al.}{2014}]{anders14}
{Anders}, F., {Chiappini}, C., {Santiago} et al.: 2014,
\newblock {\em \aap} {\bf 564}, A115

\bibitem[\protect\astroncite{{Aumer} et~al.}{2016}]{aumer16}
{Aumer}, M., {Binney}, J., and {Sch{\"o}nrich}, R.: 2016,
\newblock {\em \mnras}

\bibitem[\protect\astroncite{{Bensby} et~al.}{2011}]{bensby11}
{Bensby}, T., {Alves-Brito}, A., {Oey}, M.~S., {Yong}, D., and {Mel{\'e}ndez},
  J.: 2011,
\newblock {\em \apjl} {\bf 735}, L46

\bibitem[\protect\astroncite{{Bird} et~al.}{2013}]{bird13}
{Bird}, J.~C., {Kazantzidis}, S., {Weinberg}, D.~H., {Guedes}, J., {Callegari},
  S., {Mayer}, L., and {Madau}, P.: 2013,
\newblock {\em \apj} {\bf 773}, 43

\bibitem[\protect\astroncite{{Boeche} et~al.}{2013}]{boeche13b}
{Boeche}, C., {Siebert}, A., {Piffl}, T., {Just}, A., {Steinmetz}, M.,
  {Sharma}, S., {Kordopatis}, G., {Gilmore}, G., {Chiappini}, C., {Williams},
  M., {Grebel}, E.~K., {Bland-Hawthorn}, J., {Gibson}, B.~K., {Munari}, U.,
  {Siviero}, A., {Bienaym{\'e}}, O., {Navarro}, J.~F., {Parker}, Q.~A., {Reid},
  W., {Seabroke}, G.~M., {Watson}, F.~G., {Wyse}, R.~F.~G., and {Zwitter}, T.:
  2013,
\newblock {\em \aap} {\bf 559}, A59

\bibitem[\protect\astroncite{{Bournaud} et~al.}{2009}]{bournaud09}
{Bournaud}, F., {Elmegreen}, B.~G., and {Martig}, M.: 2009,
\newblock {\em \apjl} {\bf 707}, L1

\bibitem[\protect\astroncite{{Bovy} et~al.}{2012}]{bovy12a}
{Bovy}, J., {Rix}, H.-W., {Liu}, C., {Hogg}, D.~W., {Beers}, T.~C., and {Lee},
  Y.~S.: 2012,
\newblock {\em \apj} {\bf 753}, 148

\bibitem[\protect\astroncite{{Bovy} et~al.}{2016}]{bovy16}
{Bovy}, J., {Rix}, H.-W., {Schlafly}, E.~F., {Nidever}, D.~L., {Holtzman},
  J.~A., {Shetrone}, M., and {Beers}, T.~C.: 2016,
\newblock {\em \apj} {\bf 823}, 30

\bibitem[\protect\astroncite{{Brook} et~al.}{2005}]{brook05}
{Brook}, C.~B., {Gibson}, B.~K., {Martel}, H., and {Kawata}, D.: 2005,
\newblock {\em \apj} {\bf 630}, 298

\bibitem[\protect\astroncite{{Brook} et~al.}{2004}]{brook04}
{Brook}, C.~B., {Kawata}, D., {Gibson}, B.~K., and {Freeman}, K.~C.: 2004,
\newblock {\em \apj} {\bf 612}, 894

\bibitem[\protect\astroncite{{Brook} et~al.}{2012}]{brook12}
{Brook}, C.~B., {Stinson}, G.~S., {Gibson}, B.~K., {Kawata}, D., {House},
  E.~L., {Miranda}, M.~S., {Macci{\`o}}, A.~V., {Pilkington}, K., {Ro{\v
  s}kar}, R., {Wadsley}, J., and {Quinn}, T.~R.: 2012,
\newblock {\em \mnras} {\bf 426}, 690

\bibitem[\protect\astroncite{{Brunetti} et~al.}{2011}]{brunetti11}
{Brunetti}, M., {Chiappini}, C., and {Pfenniger}, D.: 2011,
\newblock {\em \aap} {\bf 534}, A75

\bibitem[\protect\astroncite{{Cheng} et~al.}{2012}]{cheng12a}
{Cheng}, J.~Y., {Rockosi}, C.~M., {Morrison}, H.~L., {Sch{\"o}nrich}, R.~A.,
  {Lee}, Y.~S., {Beers}, T.~C., {Bizyaev}, D., {Pan}, K., and {Schneider},
  D.~P.: 2012,
\newblock {\em \apj} {\bf 746}, 149

\bibitem[\protect\astroncite{{Chiappini}}{2009}]{chiappini09a}
{Chiappini}, C.: 2009,
\newblock in {J.~Andersen, J.~Bland-Hawthorn, \& B.~Nordstr{\"o}m} (ed.), {\em
  IAU Symposium}, Vol. 254 of {\em IAU Symposium}, pp 191--196

\bibitem[\protect\astroncite{{Comer{\'o}n} et~al.}{2012}]{comeron12}
{Comer{\'o}n}, S., {Elmegreen}, B.~G., {Salo}, H., {Laurikainen}, E.,
  {Athanassoula}, E., {Bosma}, A., {Knapen}, J.~H., {Gadotti}, D.~A., {Sheth},
  K., {Hinz}, J.~L., {Regan}, M.~W., {Gil de Paz}, A., {Mu{\~n}oz-Mateos},
  J.~C., {Men{\'e}ndez-Delmestre}, K., {Seibert}, M., {Kim}, T., {Mizusawa},
  T., {Laine}, J., {Ho}, L.~C., and {Holwerda}, B.: 2012,
\newblock {\em \apj} {\bf 759}, 98

\bibitem[\protect\astroncite{{Forbes} et~al.}{2012}]{forbes12}
{Forbes}, J., {Krumholz}, M., and {Burkert}, A.: 2012,
\newblock {\em \apj} {\bf 754}, 48

\bibitem[\protect\astroncite{{Gilmore} and {Reid}}{1983}]{gilmore83}
{Gilmore}, G. and {Reid}, N.: 1983,
\newblock {\em \mnras} {\bf 202}, 1025

\bibitem[Girardi(2016)]{girardi16} 
Girardi, L.\ 2016, \araa, 54, 95 

\bibitem[\protect\astroncite{{Grand} et~al.}{2016}]{grand16}
{Grand}, R.~J.~J., {Springel}, V., {G{\'o}mez}, F.~A., {Marinacci}, F.,
  {Pakmor}, R., {Campbell}, D.~J.~R., and {Jenkins}, A.: 2016,
\newblock {\em \mnras} {\bf 459}, 199

\bibitem[\protect\astroncite{{Juri{\'c}} et~al.}{2008}]{juric08}
{Juri{\'c}}, M., {Ivezi{\'c}}, {\v Z}., {Brooks}, A., {Lupton}, R.~H.,
  {Schlegel}, D., {Finkbeiner}, D., {Padmanabhan}, N., {Bond}, N., {Sesar}, B.,
  {Rockosi}, C.~M., {Knapp}, G.~R., {Gunn}, J.~E., {Sumi}, T., {Schneider},
  D.~P., {Barentine}, J.~C., {Brewington}, H.~J., {Brinkmann}, J., {Fukugita},
  M., {Harvanek}, M., {Kleinman}, S.~J., {Krzesinski}, J., {Long}, D.,
  {Neilsen}, Jr., E.~H., {Nitta}, A., {Snedden}, S.~A., and {York}, D.~G.:
  2008,
\newblock {\em \apj} {\bf 673}, 864

\bibitem[\protect\astroncite{{Kawata} et~al.}{2016}]{kawata16}
{Kawata}, D., {Grand}, R.~J.~J., {Gibson}, B.~K., {Casagrande}, L., {Hunt},
  J.~A.~S., and {Brook}, C.~B.: 2016,
\newblock {\em ArXiv e-prints}

\bibitem[\protect\astroncite{{Loebman} et~al.}{2011}]{loebman11}
{Loebman}, S.~R., {Ro{\v s}kar}, R., {Debattista}, V.~P., {Ivezi{\'c}}, {\v
  Z}., {Quinn}, T.~R., and {Wadsley}, J.: 2011,
\newblock {\em \apj} {\bf 737}, 8

\bibitem[\protect\astroncite{{Marinacci} et~al.}{2014}]{marinacci14}
{Marinacci}, F., {Pakmor}, R., and {Springel}, V.: 2014,
\newblock {\em \mnras} {\bf 437}, 1750

\bibitem[\protect\astroncite{{Martig} et~al.}{2012}]{martig12}
{Martig}, M., {Bournaud}, F., {Croton}, D.~J., {Dekel}, A., and {Teyssier}, R.:
  2012,
\newblock {\em \apj} {\bf 756}, 26

\bibitem[\protect\astroncite{{Martig} et~al.}{2009}]{martig09}
{Martig}, M., {Bournaud}, F., {Teyssier}, R., and {Dekel}, A.: 2009,
\newblock {\em \apj} {\bf 707}, 250

\bibitem[\protect\astroncite{{Martig} et~al.}{2016b}]{martig16a}
{Martig}, M., {Fouesneau}, M., {Rix}, H.-W., {Ness}, M., {M{\'e}sz{\'a}ros},
  S., {Garc{\'{\i}}a-Hern{\'a}ndez}, D.~A., {Pinsonneault}, M., {Serenelli},
  A., {Silva Aguirre}, V., and {Zamora}, O.: 2016,
\newblock {\em \mnras} {\bf 456}, 3655

\bibitem[\protect\astroncite{{Martig} et~al.}{2016a}]{martig16b}
Martig, M., Minchev, I., Ness, M., Fouesneau, M., \& Rix, H.-W.\ 2016, arXiv:1609.01168 


\bibitem[\protect\astroncite{{Martig} et~al.}{2014}]{martig14b}
{Martig}, M., {Minchev}, I., and {Flynn}, C.: 2014,
\newblock {\em \mnras} {\bf 443}, 2452

\bibitem[\protect\astroncite{{Minchev} et~al.}{2013}]{mcm13}
{Minchev}, I., {Chiappini}, C., and {Martig}, M.: 2013,
\newblock {\em \aap} {\bf 558}, A9

\bibitem[\protect\astroncite{{Minchev} et~al.}{2014a}]{mcm14}
{Minchev}, I., {Chiappini}, C., and {Martig}, M.: 2014a,
\newblock {\em \aap} {\bf 572}, A92

\bibitem[\protect\astroncite{{Minchev} et~al.}{2016}]{minchev16a}
Minchev, I., Chiappini, C., \& Martig, M.\ 2016, Astronomische Nachrichten, {\bf 337}, 944 

\bibitem[\protect\astroncite{{Minchev} et~al.}{2014b}]{minchev14}
{Minchev}, I., {Chiappini}, C., {Martig}, M., {Steinmetz}, M., {de Jong},
  R.~S., {Boeche}, C., {Scannapieco}, C., {Zwitter}, T., {Wyse}, R.~F.~G.,
  {Binney}, J.~J., {Bland-Hawthorn}, J., {Bienayme}, O., {Famaey}, B.,
  {Gibson}, B.~K., {Grebel}, E.~K., {Gilmore}, G., {Helmi}, A., {Kordopatis},
  G., {Lee}, Y.~S., {Munari}, U., {Navarro}, J.~F., {Parker}, Q.~A., {Quillen},
  A.~C., {Reid}, W.~A., {Siebert}, A., {Siviero}, A., {Seabroke}, G., {Watson},
  F., and {Williams}, M.: 2014b,
\newblock {\em \apjl} {\bf 781}, L20

\bibitem[\protect\astroncite{{Minchev} and {Famaey}}{2010}]{mf10}
{Minchev}, I. and {Famaey}, B.: 2010,
\newblock {\em \apj} {\bf 722}, 112

\bibitem[\protect\astroncite{{Minchev} et~al.}{2011}]{minchev11b}
{Minchev}, I., {Famaey}, B., {Quillen}, A.~C., and {Dehnen}, W.: 2011,
\newblock {\em arXiv:1111.0195}

\bibitem[\protect\astroncite{{Minchev} et~al.}{2012}]{minchev12b}
{Minchev}, I., {Famaey}, B., {Quillen}, A.~C., {Dehnen}, W., {Martig}, M., and
  {Siebert}, A.: 2012,
\newblock {\em \aap} {\bf 548}, A127

\bibitem[\protect\astroncite{{Minchev} et~al.}{2015}]{minchev15}
{Minchev}, I., {Martig}, M., {Streich}, D., {Scannapieco}, C., {de Jong},
  R.~S., and {Steinmetz}, M.: 2015,
\newblock {\em \apjl} {\bf 804}, L9

\bibitem[Miranda et al.(2016)]{miranda16} 
Miranda, M.~S., Pilkington, K., Gibson, B.~K., et al.\ 2016, \aap, 587, A10 

\bibitem[\protect\astroncite{{Noels} and {Bragaglia}}{2015}]{noels15}
{Noels}, A. and {Bragaglia}, A.: 2015,
\newblock in A. {Miglio}, P. {Eggenberger}, L. {Girardi}, and J.
  {Montalb{\'a}n} (eds.), {\em Asteroseismology of Stellar Populations in the
  Milky Way}, Vol.~39 of {\em Astrophysics and Space Science Proceedings}, p.
  167

\bibitem[\protect\astroncite{{Ojha}}{2001}]{ojha01}
{Ojha}, D.~K.: 2001,
\newblock {\em \mnras} {\bf 322}, 426

\bibitem[\protect\astroncite{{Pohlen} et~al.}{2007}]{pohlen07}
{Pohlen}, M., {Zaroubi}, S., {Peletier}, R.~F., and {Dettmar}, R.-J.: 2007,
\newblock {\em \mnras} {\bf 378}, 594

\bibitem[\protect\astroncite{{Quinn} et~al.}{1993}]{quinn93}
{Quinn}, P.~J., {Hernquist}, L., and {Fullagar}, D.~P.: 1993,
\newblock {\em \apj} {\bf 403}, 74

\bibitem[\protect\astroncite{{Robin} et~al.}{1996}]{robin96}
{Robin}, A.~C., {Haywood}, M., {Creze}, M., {Ojha}, D.~K., and {Bienayme}, O.:
  1996,
\newblock {\em \aap} {\bf 305}, 125

\bibitem[\protect\astroncite{{Ro{\v s}kar} et~al.}{2013}]{roskar13}
{Ro{\v s}kar}, R., {Debattista}, V.~P., and {Loebman}, S.~R.: 2013,
\newblock {\em \mnras} {\bf 433}, 976

\bibitem[\protect\astroncite{{Ro{\v s}kar} et~al.}{2011}]{roskar12}
Ro{\v s}kar, R., Debattista, V.~P., Quinn, T.~R., \& Wadsley, J.\ 2012, \mnras, 426, 2089 

\bibitem[\protect\astroncite{{Sch{\"o}nrich} and {Binney}}{2009}]{schonrich09b}
{Sch{\"o}nrich}, R. and {Binney}, J.: 2009,
\newblock {\em \mnras} {\bf 399}, 1145

\bibitem[\protect\astroncite{{Solway} et~al.}{2012}]{solway12}
{Solway}, M., {Sellwood}, J.~A., and {Sch{\"o}nrich}, R.: 2012,
\newblock {\em \mnras} {\bf 422}, 1363

\bibitem[\protect\astroncite{{Stinson} et~al.}{2013}]{stinson13}
{Stinson}, G.~S., {Bovy}, J., {Rix}, H.-W., {Brook}, C., {Ro{\v s}kar}, R.,
  {Dalcanton}, J.~J., {Macci{\`o}}, A.~V., {Wadsley}, J., {Couchman}, H.~M.~P.,
  and {Quinn}, T.~R.: 2013,
\newblock {\em \mnras}

\bibitem[\protect\astroncite{{Vera-Ciro} et~al.}{2014}]{vera-ciro14}
{Vera-Ciro}, C., {D'Onghia}, E., {Navarro}, J., and {Abadi}, M.: 2014,
\newblock {\em \apj} {\bf 794}, 173

\bibitem[\protect\astroncite{{Villalobos} and {Helmi}}{2008}]{villalobos08}
{Villalobos}, {\'A}. and {Helmi}, A.: 2008,
\newblock {\em \mnras} {\bf 391}, 1806

\bibitem[\protect\astroncite{{Yoachim} and {Dalcanton}}{2006}]{yoachim06}
{Yoachim}, P. and {Dalcanton}, J.~J.: 2006,
\newblock {\em \aj} {\bf 131}, 226

\bibitem[\protect\astroncite{{Yoshii}}{1982}]{yoshii82}
{Yoshii}, Y.: 1982,
\newblock {\em \pasj} {\bf 34}, 365

\end{thebibliography}
\end{document}